\documentstyle[aps,prd,epsfig]{revtex} 
\draft

\begin{document}
%&@twocolumnfalse\endcsname
\title{%
\hbox to\hsize{\normalsize\rm May 2002
\hfil Preprint MPI-PTh/2002-22}
\vskip 36pt
             Classical Radiation of a Finite Number of Photons}
\author{L.~Stodolsky}
\address{Max-Planck-Institut f\"ur Physik 
(Werner-Heisenberg-Institut),
F\"ohringer Ring 6, 80805 M\"unchen, Germany  
~~~~~~~~~~~~~~~~~~~~~~~~~~ email: les@mppmu.mpg.de}
%\date{ }
\maketitle
\bigskip
\centerline{ For {\it Acta Polonica}, ``In Honour  of the 60th
Birthday of Stefan Pokorski"}
\bigskip

\begin{abstract} Under certain conditions the number of photons
radiated classically by a charged particle following a prescribed
trajectory can be finite. An  interesting formula for this
number is presented and discussed. 

\end{abstract}
%\pacs{PACS numbers: }
\vskip2.0pc

%%%%%%%%%%%%%%%%%%%%%%%%%%%%%%%%%%%%%%%%%%%%
%%%%%%%%%%%%%%%%%%%%%
%%%%%
%% Main Text
%%%%%%%%%%%%%%%%%%%%%%%%%%%%%%%%%%%%%%%%%%%%
%%%%%%%%%%%%%
%%%%%%%%%%%%%%%%%%%%%%%%%%%%%%%%%%%%%%%%%%%%
%%%%%%%%%%%%%%%%%%%%%
%%%%%
\section{finite number of photons and their formula}
 In the classical theory of  radiation by a charged particle
 one  calculates the
energy radiated into the
electromagnetic field. Indeed from the purely classical point of
view the energy is practically the only quantity there {\it is} to
calculate. However, in certain
problems~\cite{chg} one may come upon
the idea of finding the {\it number} of photons radiated by a
charged particle following a
given trajectory. Although the photon is a quantum concept and so
the question of finding the number  might be thought to involve
quantum mechanics, it does
so only in the most minimal way. Given the energy radiated into a
given mode of the field, it is only necessary to use Planck's
relation and  divide by the frequency $\omega $ to find the number
$n$; thus the problem remains an essentially classical one.

 It might be objected that $n$ can be infinite while the energy is
finite, as in the well-known ``infrared catastrophe''. True, but as
we shall see, there is an interesting
class of cases where  this is {\it not} the case. In particular
when the trajectory of a charged particle begins and ends with the
same vector velocity $\bf v$, $n$ is generally finite. This is
because
the ``infrared catastrophe'' results from the difference in the
long flight paths for the in-and-outgoing particles, and when they
are the same, the ``catastrophe'' is averted.

 On the other hand, bremsstrahlung
calculations often have an ``ultraviolet'' or  high frequency
divergence, due to the sudden appearance or deflection of a charge
\cite{jack}. Evidently one
cannot emit an infinite number of {\it finite} energy photons, so
this divergence must be  an artifact of the calculation. This
difficulty can have either a quantum or classical
resolution. In the quantum solution, as in the feynman graph
technique, one takes into account the  energy-momentum conservation
   usually neglected at the classical level, and a suppression 
results 
for high energy photons. 
However, and more of  interest to us here, there can also be a
classical resolution: the  ultraviolet divergence results from 
abrupt
 changes in the trajectory, and  if the velocity changes or
accelerations  are sufficiently smooth, there is no divergence.

Therefore we have the interesting situation that for a smooth
trajectory, beginning and ending with the same
velocities,   the number of photons $n$ radiated
according to a classical calculation should be finite.
But  $n$ is  a dimensionless,  lorentz-invariant quantity. Thus
there
ought to be some simple formula for  it, a relation
mapping the path of a  particle in
space-time to the real number, $n$.

This relation is:
\begin{equation} \label{n} 
n = {\alpha\over \pi } \int \int dx_{\mu}{1\over S^2_{i\epsilon}}
dx'_{\mu}
\end{equation}
In this formula $x$ and $x'$ are four-dimensional coordinates,
referring to points on the space-time path of the charged particle,
so that $dx_{\mu}$ is a 4-vectorial element of the path  (Fig 1).
One may  introduce the proper time or invariant path length $\tau$
and the  four-velocity $u_\mu =dx_\mu/d\tau$ to also write
the expression as
\begin{equation} \label{nn} 
n = {\alpha\over \pi }\int \int {dx_{\mu}\over d\tau}{1\over
S^2_{i\epsilon}}
{dx_{\mu}\over d\tau'}~ d\tau d\tau'= {\alpha\over \pi }\int \int
u_\mu(\tau){1\over S^2_{i\epsilon}}~u_\mu(\tau')~ d\tau d\tau'
\end{equation}

$S_{i\epsilon}$
is the four-distance between the points $x,x'$ in the following way
\begin{equation} \label{s} 
S^2_{i\epsilon}= (t-t'+i\epsilon)^2 -({\bf x-x'})^2 
\end{equation}
As we shall explain shortly the $i\epsilon$ is necessary to make
sense of and to properly define~Eq~[\ref{n}].

 It is interesting  
that the Sommerfeld fine structure constant $\alpha= {e^2\over
\hbar c}
\approx 1/137$, involving
Planck's quantum constant, appears. It arises, however, not
 as a coupling constant, but rather from the conversion of the
classical
energy to quanta, where we finally must
divide by $\hbar \omega$. The  expression is a non-perturbative.
If the particle is multiply charged with $Q$
electron charges, the formula should be multiplied by $Q^2$.
Observe that by interchanging $t$ and $t'$ in the integrations one
can
reverse the sign of $i\epsilon$  to show that $n$ is real.
The expression is not an integer since it represents the average or
expectation value of the number of particles radiated.
 
\begin{figure}[h]
\epsfig{file=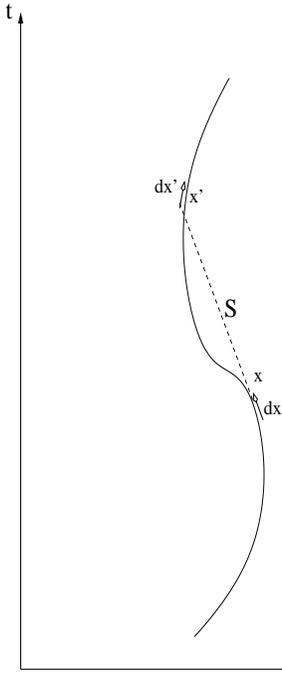, height=.5\hsize, width=.5\hsize}
\caption{Path in space-time}
\end{figure}
\section{derivation}
We can derive~Eq~[\ref{n}] from the standard
treatments of classical radiation theory~\cite{jack},~\cite{ll}
where one
calculates
${\bf j}({\bf k})$, the fourier
components of the three-vector current, and then finds the energy
radiation rate $\sim |{\bf j}|^2 sin^2\theta$, where $\theta$ is
the angle between ${\bf k}$ and ${\bf j}$. Dividing by  $\omega$
(units $\hbar=1$)gives $n$. 
 
To obtain the total number,
we  sum over all modes ${\bf k}$ of the radiation field,
\begin{equation}\label{nj}
n\sim\int  sin^2\theta |{\bf j}({\bf k})|^2{1\over \omega}~ d^3k\;.
\end{equation}

  One could  try to perform the integrations for every
particular trajectory, but we would
like to find a general formula in terms of the path itself.  Thus
we attempt to perform the $d^3k$ integration first,
before
the fourier transform. This gives at first a not well-defined
expression, which leads to the need to introduce the
$i\epsilon$.

To see how this comes about, we observe that for smooth paths ${\bf
x}(t)$ the fourier transform ${ \bf j}({\bf k})$
is strongly convergent at large $\omega$. 
When we introduce
the current density for our classical point particle as

\begin{equation}\label{cur}
{\bf j}(t,{\bf x})= e {\bf v}(t)\delta ^3({\bf x}-{\bf x}(t))
\end{equation}
 we have  ${\bf j}({\bf k})=e\int dt\, {\bf v}e^{i({\bf
kx}(t)-\omega
t)}$
for the fourier transform. Since the path of the particle is
 time--like the exponent can never become zero, and if $\bf x$ or
${\bf
v}={d \over dt} {\bf x}$ have no jumps or kinks the integral
vanishes
rapidly for large $\omega$.  Eq~[\ref{nj}] can be rewritten in
a more covariant way using $sin^2\theta= 1-cos^2\theta$ and noting
the current conservation relation $\omega
j_0=\omega cos\theta \vert {\bf j}\vert  $, so that
Eq~[\ref{nj}] is essentially the integral of the four-vector
squared ${\bf j}^2-j_0^2=-
j_{\mu}^2$.
Using $dt~ {\bf v}= d{\bf x}$
 one has a compact line integral
expression for $j_{\mu}$, namely
\begin{equation}\label{jmu}
 j_{\mu}({\bf k})= e \int dx_{\mu}e^{ikx}\;.
\end{equation}
where $kx$ is the four dimensional scalar product
$k_{\nu}x_{\nu}=k_0t-{\bf kx}(t)$, with $k_0=\omega=|{\bf k}|$

Squaring Eq~[\ref{jmu}] and writing in the constants~\cite{ll} we
get for the integral of $-
j_{\mu}^2(k)$ over all modes of the radiation field:

\begin{equation}\label{jmua}
n= -{\alpha\over 4\pi^2 } \int d^3k {1\over \omega} \int\int
dx_{\mu}
dx'_{\mu}e^{ik(x-
x')}\;,
\end{equation}
 Firm in the belief that we actually have a well defined,
convergent expression, we allow ourselves to do
the $d^3k$ integral first and to use some $i\epsilon$ manipulations
in dealing with seemingly ill-defined expressions.

 After a few steps we
get to the
expression $\int^{\infty}_0 \omega\em d\omega\em e^{i\omega((t-t')-
R cos\theta )}$, where $R=|\bf{x-x'}|$ and where here $\theta$ is
the angle
between
$\bf k$ and the vector $(\bf{x-x'})$. Now, since we believe that
our expressions are rapidly vanishing at large $\omega$ and
$\omega$ is always positive, it shouldn't hurt to
add an $i\epsilon$ in the exponent~\cite{gelf}, to make the
replacement
$(t-t')\rightarrow (t-t'+i\epsilon)$.
The integral is then convergent, and
integrating over $d\em cos \theta$ and keeping track of the $i
\epsilon$,  this leads to $\int d^3k {1\over \omega}e^{ik(x-
x')}=-4\pi {1\over S^2_{i\epsilon}}$
 and so Eq~[\ref{n}].

Although these manipulations are similar to those with
the invariant D functions in {\it quantum} field theory, and our
function $1/S^2$ resembles a $D$ function, our assumption of
smooth paths has lead us to an ``$i\epsilon$ prescription'' in
position rather than in momentum space.

 \section{Elimination of $\epsilon$}
 The difficult job now becomes the evaluation of the $i\epsilon$ in
Eq~[\ref{n}]. We look for guidance to the theory of generalized
functions~\cite{gelf} where one has the relation 

\begin{equation}\label{f0}
\int_{-\infty}^{+\infty} {f(t)\over (t+i\epsilon)^2}dt=\int_{-
\infty}^{+\infty} {f(t)-f(0)\over t^2} dt\; ,
\end{equation}
where we have specialized to the case of an even function
$f(t)=f(-t)$.

 One way of understanding this relation  is to note that, for
$\epsilon$ non-zero, 

\begin{equation}\label{f1}
 \int_{-\infty}^{+\infty} {1\over
(t+i\epsilon)^2}dt=0 \; ,
\end{equation}
which just follows from explicit integration.
 Thus we have simply  subtracted zero in Eq~[\ref{f0}], and have
chosen the coefficient of this zero term
in such a way as to cancel the singularity of the integrand. 
With $f$ even, $f(t)-f(0)\sim t^2$, and the expression is indeed
finite at the singularity.

 Can we apply the same idea here? That is, by examining the
neighborhood of the singularity for small but non-zero $\epsilon$
we see that our expression is finite and $\epsilon$
independent.
We would thus like to subtract
something which is zero for finite $\epsilon$ and which will
regulate the singularity at $t=t'$ in Eq~[\ref{n}], enabling us to
finally dispense with $\epsilon$ altogether.   The problem,
however,  would
seem to be much
more difficult  than in Eq~[\ref{f0}]. There we simply had to
adjust one constant, namely $f(0)$. Here we need a
function $f(t,t')$, such that when it is integrated over
$1/S^2(t,t')$
gives
zero and then can be adjusted to cancel the numerator $dx_\mu
dx'_\mu$ when
we
make the replacement $dx_\mu dx'_\mu \rightarrow dx_\mu
dx'_\mu(1-f)$. Furthermore
it should do this  for {\it any } path ${\bf x}(t)$ we care to  put
in the formula.

 This last requirement rules out, for example, the at first likely-
looking candidate replacement  $dx_\mu dx'_\mu=u_{\mu}u'_{\mu}d\tau
d\tau'\rightarrow (u_{\mu}u'_{\mu}-1)d\tau d\tau' $. Since
$u_{\mu}u'_{\mu}\rightarrow 1$ for $\tau
\rightarrow \tau'$ this would regulate the singularity and looks
promising. But it doesn't work, $\int \int d\tau d\tau'
~1/S^2(\tau,\tau')$ is obviously path dependent and cannot be zero
for
all paths.

 This, however, suggests a solution. If we can  find something for
$f$ which is a total derivative and so can be  integrated,  it
would depend only  on the end points of the path and not on the
path itself. 
 To this end,   note the following useful relation. Let $G$ be
some function
of $S^2(\tau,\tau')=\Delta _{\mu}^2$ where $\Delta
_{\mu}=x_{\mu}(\tau)-
x_{\mu}(\tau')$, then
\begin{equation} \label{tautau}
\partial_{\tau}\partial_{\tau'} G(S^2)=-G''4 (\Delta_{\mu}
{dx_{\mu}\over d\tau})(\Delta_{\nu} {dx_{\nu}\over d\tau'})
- 2G'{dx_{\mu}\over d\tau}{ dx_{\mu}\over d\tau'}\; ,
\end{equation}
where $G'$ and $G''$ refer to first and second derivatives with
respect to $S^2$.
 
Since the expression is a total derivative and its  integral is
independent of the path and  is to be interpreted as zero, we may
look for some choice for $G$ that leads to $\sim 1/S^2$ for
$\tau\rightarrow\tau'$ and so might
cancel the singularity. This  occurs in fact if we consider $G=ln
S^2$.  The above equation becomes

\begin{equation} \label{tautaua}
\partial_{\tau}\partial_{\tau'} ln(S^2)= +{4\over S^4}(\Delta
_{\mu} {dx_{\mu}\over d\tau})(\Delta _{\nu} {dx_{\nu}\over
d\tau'})- {2\over S^2} {dx_{\mu}\over d\tau}{ dx_{\mu}\over
d\tau'}\; .
\end{equation}
 Since  ${\Delta_{\mu}\over S}\rightarrow
{dx_{\mu}\over d\tau}$ for $\tau\rightarrow\tau'$ the expression
$\sim 1/S^2$ at the singularity.
Now we want to add this  to Eq~[\ref{nn}], that  is to $\int
{dx_{\mu}\over d\tau}{dx_{\mu}\over d\tau'}(1/S^2)  d\tau d\tau'$,
with some weight such that the singularity at $\tau=\tau'$ is
cancelled. To determine this weight  note that since  the four
velocity
satisfies $({dx_{\mu}\over d\tau})^2=1$  the rhs of
Eq~[\ref{tautaua}] goes to $+2/S^2$ for $\tau\rightarrow\tau'$.
Hence we must subtract  $1/2$ of Eq~[\ref{tautaua}] from
Eq~[\ref{nn}] to remove the singularity. Doing this we end up with
the following nice relation.

\begin{equation} \label{naa} 
 \int \int {dx_{\mu}\over d\tau }{1\over S_{i\epsilon}^2}
{dx_{\mu}\over d\tau'}~d\tau d\tau'= 2 \int \int{dx_{\mu}\over
d\tau }{\delta_{\mu \nu}- {\Delta_\mu \Delta_\nu\over S^2 }
\over S^2} {dx_{\nu}\over d\tau'}~d\tau d\tau'
\end{equation}
or alternatively
\begin{equation} \label{naab} 
 \int \int dx_{\mu}{1\over S_{i\epsilon}^2}
dx'_{\mu}=2 \int \int dx_{\mu} {\delta_{\mu \nu}- {\Delta_\mu
\Delta_\nu\over S^2 }
\over S^2} dx'_{\nu}
\end{equation}
The notation is meant to indicate that while the $i\epsilon$ is in
$S^2$ on the left, it is not needed on the right.

We can rewrite these equations in an interesting way by noting that
$\Delta_{\mu}/S$ is like an average 4-velocity connecting the
points $\tau,\tau'$, which we might call
$ U_{\mu}$ in analogy to the instantaneous 4-velocity
$u_{\mu}={dx_{\mu}\over
d\tau }$.

\begin{equation}\label{str}
U_{\mu}(\tau,\tau') ={\Delta_\mu
\over S}={x _{\mu}(\tau)-x _{\mu}(\tau')\over S}
\end{equation}

Then we can write

\begin{equation} \label{naau} 
 n={\alpha\over \pi }2 \int \int u_{\mu} {\delta_{\mu \nu}-
{\Delta_\mu \Delta_\nu\over
S^2 }
\over S^2} u_{\nu}~d\tau d\tau'= {\alpha\over \pi }2\int \int
{u(\tau)\cdot u(\tau')- \bigr(U\cdot u(\tau)\bigl) \bigr(U\cdot
u(\tau')\bigl)
\over S^2}~d\tau d\tau'
\end{equation}

We can now introduce the
notion of a ``transverse vector'' $u^T$ associated with the points
$x,x'$ at $\tau, \tau'$. This is the  vector $u$   with the
``longitudinal'' part,
that is the component
along $U$,  removed:

\begin{equation} \label{udef}
u_\mu^T(\tau,\tau')=u_\mu(\tau)- U_\mu \bigr(U\cdot
u(\tau)\bigl)~~~~~~~~~~~~~~u_\mu^T(\tau',\tau)=u_\mu(\tau')- U_\mu
\bigr(U\cdot u(\tau')\bigl)
\end{equation}
 With these
definitions,
$u_\mu^T(x,x')u_\mu^T(x',x)=u_\mu(x)(\delta _{\mu \nu}- U_\mu
U_\nu) u_\nu (x')$, which is the numerator in 
Eq~[\ref{naau}] and we can write the above
equations in terms of products of ``transverse vectors'':
\begin{equation} \label{nabu}
n= {\alpha\over \pi }~ 2 \int \int u_\mu^T(\tau,\tau')
{1\over S^2} u_\mu^T(\tau',\tau)  ~d\tau d\tau'\; .
\end{equation}
 
Hence we can say that our integral for $n$ represents the
``interaction'' of pairs of transverse vectors along the path of
the
charge. The possible singularity at $S=0$ is absent because
$u^T(\tau,\tau')$
vanishes for two points  very close together.

\section{Properties of the quantities}
We summarize some properties of these quantities: $U$ and $u$ are
unit vectors, $u^2=U^2=1$. 
From their definitions $U$ and $u^T$ are orthogonal
$u^T_\mu(\tau,\tau') U_\mu=u^T_\mu(\tau',\tau) U_\mu=0$.
 Note for straight line
motion or more generally for any smooth path as
$\tau\rightarrow\tau'$

\begin{equation} \label{uu}
U_\mu\rightarrow u_\mu
\end{equation}
Unlike $u$ and $U$, $u^T$ is a space-like
(or zero) vector. This follows from the fact that there is a frame
where the time component of $u^T$ vanishes, namely the restframe of
the time-like $U$.

 We note a point concerning the definition of $U$  in
Eq~[\ref{str}]. This point is irrelevant in all those expressions
where $S$ or $U$ appears quadratically, but we mention it for
consistency.
 Namely, we would like $U$ to resemble the velocity. Therefore it
must be understood that $S(\tau,\tau')=\sqrt{S^2}$ is ``directed'';
an odd
function,
positive
for $\tau >\tau'$ and negative for $\tau <\tau'$, like $(\tau -
\tau')$.  It implies a definition of $U$ such that
$U(\tau,\tau')=U(\tau',\tau)$ and that $U$ is ``forward pointing'',
i.e. $U_0$ is always positive.

 An interesting expression for $u^T$ follows
from the definition of $U$ in Eq~[\ref{str}], reflecting the fact
that the derivative of a unit vector is transverse to itself:

\begin{equation} \label{ut}
{1\over
S}~u^T_\mu
(\tau,\tau')=\partial_{\tau}U_\mu(\tau,\tau')~~~~~~~~~~~~~~~~~~~
-{1\over S}~u^T_\mu(\tau',\tau)=\partial_{\tau'}U_\mu(\tau,\tau')
\end{equation}

With this, we can also write Eq~[\ref{nabu}] in some different
looking ways:
\begin{equation} \label{aa}
n= -{\alpha\over \pi } ~2\int \int             
\partial_{\tau}U_{\mu}(\tau,\tau')~~\partial_{\tau'}U_{\mu}(\tau, 
\tau') ~d\tau d\tau'
\end{equation}

Or, since $U^2=1$  and so 
$\partial_{\tau}\partial_{\tau'}(U_\mu U_\mu)=0$ we can also write

\begin{equation} \label{aaa}
n= {\alpha\over \pi } ~2\int \int             
U_{\mu}(\tau,\tau')~\partial_{\tau}~\partial_{\tau'}U_{\mu}(\tau, 
\tau') ~d\tau d\tau'
\end{equation}

Another variant results if we note that~Eq~[\ref{aa}] looks like
the cross terms of a quadratic expression:
$2\partial_{\tau}U\partial_{\tau'}U= {1\over 2}
[(\partial_{\tau}U+\partial_{\tau'}U)^2-(\partial_{\tau}U-
\partial_{\tau'}U)^2]$. Introducing the sum and difference
variables $\tau_+={1\over 2}(\tau+\tau')$ and $\tau_-={1\over
2}(\tau'-\tau)$ 
    
\begin{equation} \label{aab}
n= {\alpha\over \pi } ~\int \int             
~[(\partial_{\tau_-}U)^2 -~(\partial_{\tau_+}U)^2 ]~d\tau_+ d\tau_-
\; ,
\end{equation}
where  because of the rotation of the coordinates in the $\tau,
\tau'$ plane, the integration  boundaries now have the form of a
diamond instead of a square. 

\section{Infrared Behavior}
 The $i\epsilon$ takes care of the high frequency or ultraviolet
behavior, but there might be questions concerning the infrared
region. We see this very simply  if we consider the stationary
particle with ${\bf v}=0$, for which $n$ of course should be zero.
Eq[\ref{n}] leads to

\begin{equation} \label{stat}
n=\int_{-\infty}^{+\infty} \int_{-\infty}^{+\infty} dt dt'{1\over
(t-t' +i\epsilon)^2}
\end{equation}
 There are two kinds of limits implied here $\epsilon\rightarrow
0$, and some upper/lower limit of the integration $L\rightarrow
\infty$, and the  integral can depend on how the limits are taken.
If
we simply apply Eq~[\ref{f1}] we get of course zero, as desired. On
the other hand if we interpret the limits at $\pm \infty$ by, say,
integrating over some test function which is constant up to some
large number $L$ and then drops off, we can get an answer involving
$\epsilon L$, which  depends on the order in which we take
$\epsilon\rightarrow 0$, $L\rightarrow \infty$. The problem
originates in the fact that we are turning off the charge at some
$L$,  violating  charge conservation and in the process producing
some photons. We are thus to handle this limit by remembering that
$\epsilon$ is to be kept finite to the very end, and so we first
mean $\epsilon L\rightarrow \infty$. Alternatively one can subtract
the integral for the stationary particle, as in  Eq~[\ref{stat}] to
regularize the infrared behavior.

 Naturally if we use the form without the $i\epsilon$, say from 
Eq~[\ref{naau}] we immediately
get zero for the stationary particle. More generally, we can
investigate the contribution of the final and initial long straight
paths for a moving particle
in Eq~[\ref{naau}]. 
 The dangerous regions for the infrared behavior are large
$\tau,\tau'$ as the particle comes from or goes to infinity. For
both $\tau$'s very large or both very small, we have in view of
Eq~[\ref{uu}] and $u^2=1$ that  the numerator becomes $u^2-
u^2u^2=0$. Similarly if one $\tau$ is at very early times and the
other a very late times and the velocities at these times are the
same, there is the analogy to Eq~[\ref{uu}]

\begin{equation}\label{stra}
U_\mu(\tau,\tau')=u_\mu+ {\cal O}({T\over \tau-\tau'})
\end{equation}
where $T$ is the  finite time period where the charge was not in
uniform motion. For large $(\tau-\tau')$, $U\rightarrow u$ and the
numerator will again tend to zero, as long as the initial and final
velocities are the same. It is interesting that the elimination of
both the long and short distance singularities can in a sense be
attributed to the same relation, Eq~[\ref{uu}].

\section{Non-relativistic limit}

Consider that class of paths  whose tangents are  roughly
parallel to some straight line on Fig (1), meaning that the
ordinary
velocity $\bf v$ is always close to some typical or average
velocity. By
making a lorentz transformation (under which $n$ is invariant)
 so that this typical velocity is zero, we have as a first
approximation in the ordinary velocity   the
non-relativistic limit, leading to expressions  
quadratic in velocities. 

We could make a stab at the non-relativistic limit by taking our
basic expression Eq~[\ref{n}] and just naively expanding it for
small velocities. We introduce $\bf V$, the average velocity
vector 
connecting two points on the  curve 
\begin{equation} \label{v}
 {\bf V}={{\bf x}(t)-{\bf x}(t')\over t-t'}\;,
\end{equation}
 ${\bf V}$ is a symmetric quantity ${\bf V}(t,t')={\bf V}(t',t)$,
in parallel with our earlier definition of $U$ and
becomes equal to $\bf v$ when  $t\rightarrow t'$. ${\bf V}$ plays
a role analogous to $U$ except that it does not have a fixed length
and so $V^2$ has non-zero time derivatives. 
Expanding in Eq~[\ref{n}] we can try to write

\begin{equation}\label{sex} 
1/S^2_{i\epsilon}\approx {1\over (t-t'+i\epsilon)^2}~(1+
V^2_{i\epsilon}(t,t')+V^4_{i\epsilon}(t,t')+...)
\end{equation}

 where ${\bf
V}_{i\epsilon}={({\bf x}-{\bf x'})\over (t-t'+i\epsilon)}$
so that 
\begin{equation} \label{nna} 
n ={\alpha\over \pi} \int \int
dx_{\mu} {1\over S^2_{i\epsilon}}dx'_{\mu}\approx {\alpha\over \pi}
\int \int
dt dt'(1-{\bf v}(t){\bf v}(t')){1\over (t-t'+i\epsilon)^2}~(1+
V^2_{i\epsilon}(t,t')+V^4_{i\epsilon}(t,t')+...)~\; .
\end{equation}
 
Now,  using Eq~[\ref{f1}] the ``1'' term vanishes and we have

\begin{equation} \label{another}
n= {\alpha\over \pi} \int \int dt dt'{1\over (t-t')^2}[V^2-{\bf
v}(t){\bf
v}(t')] ~~~~~~~~(non-rel.~ limit)
\end{equation}
We have dropped the $i\epsilon$ since the expression is  
non-singular  with ${\bf V}\rightarrow {\bf
v}$ as  $t\rightarrow t'$.

This is a not unreasonable-looking expression. Indeed, if we
expand it for $t\approx t'$ we get

\begin{equation} \label{asq}
[V^2-{\bf v}(t){\bf v}(t')]\approx {1\over 4} a^2 (t-t')^2 ...\; , 
\end{equation}
showing that the singularity is cancelled and that the leading
terms
of the expression are positive and exhibit the familiar connection
between acceleration squared and radiation.
The (...) includes terms involving time derivatives of the
acceleration $a$ (${\bf a} ={d {\bf v}\over dt} $) as well as
higher order
terms in $(t-t')^2$.

One might have some qualms about the carefree $i\epsilon$
manipulations, and we indicate how to arrive
 at Eq~[\ref{another}] by straightforward application of our more
conventional formulas.

First note
 that  $U$ can be written in terms of $V$ in the usual way
relating a three-velocity and a four-velocity, $U_0=1/\sqrt{1-
V^2},~~{\bf U}={\bf V}/\sqrt{1-V^2}$ with the non-relativistic
limits
$U_0\approx 1-1/2 V^2,~~ {\bf U}\approx {\bf V} $.
 we find from either Eq~[\ref{aa}] or  Eq~[\ref{aaa}]  to
leading order in $V$
\begin{equation}\label{jmuc}
n= {\alpha\over \pi} 2 \int\int dt dt'\;\partial_t {\bf
V}(t,t')~\partial_{t'} {\bf V}(t,t')~~~~~~~~~~~~~~~(non-rel.~
limit)
\end{equation}

Now in analogy to Eq~[\ref{ut}] we have from the definition of $\bf
V$
\begin{equation}\label{vs}
\partial_t {\bf V}(t,t')={{\bf v}(t)-{\bf V}\over
t-t'}~~~~~~~~~\partial_{t'} {\bf V}(t,t')=-{{\bf v}(t')-{\bf
V}\over t-t'}
\end{equation}
and  Eq~[\ref{jmuc}]  becomes
\begin{equation}\label{jmub}
n= {\alpha\over \pi } 2 \int\int dt dt'[-{\bf v}(t){\bf
v}(t')+\bigl({\bf v}(t)+{\bf v}(t')\bigr){\bf
V}-V^2]{1\over
(t-t')^2}~~~~~~~~~~~~~~~(non-rel.~ limit)
\end{equation}
 One could be satisfied with this formula as it is, but to bring it
into the perhaps simpler form Eq~[\ref{another}], note the
following identity: ${1\over 2}\partial_t
\partial_{t'} 
V^2= [-{\bf v}(t){\bf v}(t')+2({\bf v}(t)+{\bf v}(t')){\bf
V}-3V^2](t-t')^{-2}$, which 
follows from differentiating $V^2=({\bf x}-{\bf x'})^2(t-t')^{-2}$
twice.
We split this into two parts:
${1\over 2}\partial_t \partial_{t'} 
V^2= [-{\bf v}(t){\bf v}(t')+({\bf v}(t)+{\bf v}(t')){\bf
V}-V^2](t-t')^{-2}+[({\bf v}(t)+{\bf v}(t')){\bf
V})-2V^2](t-t')^{-2}$, where we make the split  so that the first
part corresponds to Eq~[\ref{jmub}]. Now $ \partial_t \partial_{t'}
V^2 $ is  a total derivative whose integral  may be set to zero.
Therefore the integral of the first and second parts represents the
same quantity with opposite signs and we can write:  
\begin{equation}\label{jmuz}
n= {\alpha\over \pi } 2 \int\int dt dt'[-{\bf v}(t){\bf
v}(t')+({\bf v}(t)+{\bf v}(t')){\bf
V}-V^2]{1\over
(t-t')^2}=-{\alpha\over \pi } 2 \int\int dt dt'[({\bf v}(t)+{\bf
v}(t')){\bf
V})-2V^2]{1\over
(t-t')^2}~~~~~~~~~~~~~~~(non-rel.~ limit)
\end{equation}

Taking the one-half the sum of the two forms we  finally
obtain, after much labor, Eq~[\ref{another}]. The cavalier
$i\epsilon$ manipulations were certainly a lot quicker!

Despite the familiar acceleration squared in Eq~[\ref{asq}], we
shouldn't expect that  $n$ can be represented simply
by an integral of  some local quantity along the path. There is the
(...), and all our expressions are bi-local in the time. The photon
is a non-local concept and  a certain space-time interval
is necessary to define it.
This corresponds to the distinctive property of relativistic local
field
theory that while one has local constructions for quantities like
charge density or energy density, there is in fact no local 
quantity for
particle number or photon density. Indeed, the need to have 
some space-time interval to define a particle, leads to the concept
of the ``formation
zone''~\cite{zone},  which can be used to understand certain
phenomena like the absence of ``cascading'' for particle production
on nuclear targets.

\section{Simple cases}
With the simple
Eq~[\ref{another}] in hand we can proceed to calculate a couple of
concrete examples.

\bigskip

{\it Dipole radiation: } We first take the classic problem of
dipole radiation. Let a charged particle be oscillating in one
dimension
according to $x=x_0sin\Omega t$, so $v=x_0 \Omega cos\Omega t$.
Changing variables $\Omega t\rightarrow t$ and similarly for
$t'$, Eq~[\ref{another}] becomes

\begin{equation} \label{dip}
n={\alpha\over \pi} (x_0 \Omega)^2 \int \int dt
dt'\bigl[{(sin\,t-sin\,t')^2\over (t-t'
+i\epsilon )^4} -{cos\,t~
cos\,t'\over (t-t' +i\epsilon)^2} \bigr]
\end{equation}
Although the $i\epsilon$ is not necessary since the combination of
the two terms gives something non-singular, it
is convenient to keep it since it allows us to handle each term
separately.
Carrying out, say, the $t$ integral first and using relations of
the
type $\int dt{cos\, t\over (t-t'+i\epsilon)^2 }= 2\pi e^{-i t'} $
or $\int dt{sin\,t\over (t-t'+i\epsilon)^4 }= \pi (i/6) e^{-i t'}
$ and $\int dt{cos\, t'\over (t-t'+i\epsilon)^2 }= 0 $, we
arrive at
\begin{equation} \label{dipa}
n={\alpha\over \pi} (x_0 \Omega)^2 \int  dt'  \bigl[ {-i\pi \over
3}sin\,t'-(-\pi) cos\,t'\bigr]e^{-it'}=\alpha (x_0 \Omega)^2{
T\Omega\over 3 }
\end{equation}
where $T$ is the length of time the particle is in motion.
The number of photons generated  increases linearly with time, as
was to be expected. We have neglected a contribution, not
proportional to $T$, connected with turning the motion on and off. 
 If we now
multiply by $\Omega$ to find the
energy and divide by T to get  the power, we obtain $Power=
\alpha{1\over
3} (x_0\Omega)^2\Omega^2$, which  is the classical formula for
dipole
radiation averaged over a cycle~\cite{jack},~\cite{ll}. Not
surprisingly we
recover the classical result, as was our starting point. 

 Note however, that in general the energy radiated and the number
of photons do not stand
in direct relation   since the oscillating
charge produces higher harmonics  in addition to
the fundamental at frequency $\Omega$~\cite{jack}. However,  in the
nonrelativistic limit $v/c \rightarrow 0$ these higher harmonics
become negligible (since they are a retardation effect), and we
expect the  energy radiation  to be simply
proportional to $n$.

This example shows that despite our requirement  that the initial
and final velocities be equal, the method need not be of purely
academic interest for practical calculations. If the effects
connected with turning the motion
on and off are negligible compared to some main
effect, we can always return the particle to, say, zero velocity,
while retaining  the main effect. 
 
\bigskip
{\it Smooth deflection: }
Instead of an oscillator which is on for  a long time we can
consider a charge undergoing  a smooth deflection, for example
$x=x_0 {1\over 1+(t/t_0)^2}$ so that $v=-2{x_0\over t_0}{t\over
t_0}{1\over 1+(t/t_0)^2}$. This leads to
\begin {equation} \label{defl} 
n= {\alpha\over \pi} \int \int dt dt'{1\over
(t-t'+i\epsilon)^2}[V^2-{\bf
v}(t){\bf
v}(t')]={\alpha\over \pi}({x_0\over t_0})^2({-1\over 8}\pi^2-{-
3\over 8}\pi^2)={\alpha \pi\over 4}({x_0\over t_0})^2
\end{equation}

\section{Further questions}
 Eq~[\ref{n}] should have some general  symmetry
properties with respect to changing the path. There are the 
evident invariances  under lorentz transformations,
translation, 3-D rotation,  reflection, and time-reversal. Since
there are no dimensional quantities except the path itself
involved, there is
 also an
invariance under
 rescaling of all 4-coordinates simultaneously, $x_\mu \rightarrow
\lambda x_\mu $. That is to
say, if the path is expanded in space and time proportionally so
that the velocities remain unchanged, $n$ is unchanged, as we see
in the examples. It would be
interesting to know if there are further invariances and what the
full invariance group is.

 Also, we might consider the problem
for gravitons instead of photons.  Presumably  one will find, in
analogy to
Eq~[\ref{nn}], and in view of the tensorial character of the
source of gravitons, $n\sim G m^2 \int \int
u_\mu(\tau)u_\nu(\tau){1\over S^2_{i\epsilon} }u_\mu(\tau')
\mu_\nu(\tau')~ d\tau d\tau'$ where G is the gravitational constant
and $m$ the mass of the
radiating particle (recall $\hbar,c =1$, so $G m^2=(m/M_{pl})^2$),
but it would be interesting
to investigate this more closely.

 Finally, a number of interesting
mathematical problems suggest themselves. Our $n$  gives an
invariant characterization of the ``wiggly-ness'' of a curve. There
appears to be no reason why it should  not be also used in
euclidean space where $S^2(\tau,\tau')= \Sigma \bigl(x_i(\tau)-
x_i(\tau')\bigr)^2$. The main difference would seem to be that the
curve can now go ``backwards'', opening the possibility of closed
curves. As well, there is the possibility of new singularities when
two parts of the curve, with remote values of $\tau$, come close
together.

 For a closed plane curve it is plausible that the minimum
value of $n$ obtains for the circle. Then defining
   $n=-2\int \int {1\over S^2} {\bf u}^T {\bf u}'^T d\tau
d\tau'$ ( we use Eq~[\ref{nabu}], leave
away ${\alpha\over \pi}$, and the natural sign is now minus), we
thus conjecture that the
 minimum value of $n$ for any plane curve is  $2\pi^2$, which is
what we obtain for the circle.

Also, there
may be  a ``topological'' aspect to $n$, connected with knots.  For
an open path the minimum value of $n$, namely zero, is reached for
a straight line.  If the  path has a knot, however, it  cannot be
continuously deformed to a straight line, and must  go
``backwards'' somewhere, suggesting  that the minimum 
value of $n$ continuously attainable is related to the presence of
knots.

%%%%%%%%%%%%%%%%%%%%%%%%%%%%%%%%%%%%%%%%%%%%
%%%%%%%%%%%%%%%%%%%%%
%%%%%

I am grateful to Y. Frishman, D.Maison, and V. Zakharov for many
and helpful discussions. 
%%%%%%%%%%%%%%%%%%%%%%%%%%%%%%%%%%%%%%%%%%%%
%%%%%%%
%%%%%%%%%%%%%%%%%%%%%%%%%%%%%%%%%%%%%%%%%%%%
%%%%%%%%%%%%%%%%%%%%%
%%%%%

%\section*{Acknowledgments}

%I am indebted to
%%%%%%%%%%%%%%%%%%%%%%%%%%%%%%%%%%%%%%%%%%%%
%%%%%%%%%%%%%%%%%%%%%
%%%%%
%% References
%%%%%%%%%%%%%%%%%%%%%%%%%%%%%%%%%%%%%%%%%%%%
%%%%%%%%%%%%
%%%%%%%%%%%%%%%%%%%%%%%%%%%%%%%%%%%%%%%%%%%%
%%%%%%%%%%%%%%%%%%%%%
%%%%%

\end{document}